\documentclass[aps,prl,twocolumn,showkeys,amsmath,amssymb]{revtex4}
\usepackage{graphicx}% Include figure files
\usepackage{dcolumn}% Align table columns on decimal point
\usepackage{bm}% bold math
\usepackage{subfigure}
\usepackage{placeins}

\begin{document}

\title{All-diamond optical assemblies for a beam-multiplexing X-ray monochromator at the Linac Coherent Light Source
\footnote{submitted for publication to Journal of Applied Crystallography http://journals.iucr.org/j/}}

\author{S. Stoupin$^1$, 
S.A. Terentyev$^2$, 
V.D. Blank$^2$, 
Yu.V. Shvyd'ko$^1$,
K.Goetze$^1$, 
L. Assoufid$^1$, 
S.N. Polyakov$^{2,3}$, 
M.S. Kuznetsov$^2$, 
N.V. Kornilov$^2$,
J. Katsoudas$^4$,
R. Alonso-Mori$^5$,
M. Chollet$^5$,
Y. Feng$^5$,
J.M. Glownia$^5$,
H. Lemke$^5$,
A. Robert$^5$,
S. Song$^5$,
M. Sikorski$^5$,
D. Zhu$^5$}

\affiliation{
$^1$Advanced Photon Source, Argonne National Laboratory, Argonne, IL, USA \\
$^2$Technological Institute for Superhard and Novel Carbon Materials, Troitsk, Russia \\
$^3$Skobeltsyn Institute of Nuclear Physics, Lomonosov Moscow State University, Moscow, Russia \\
$^4$Illinois Institude of Technology, Chicago, IL, USA \\
$^5$Linac Coherent Light Source, SLAC National Accelerator Laboratory, Menlo Park, CA, USA}

\begin{abstract}
A double-crystal diamond (111) monochromator recently implemented at the Linac Coherent Light Source (LCLS) 
enables splitting of the primary X-ray beam into a pink (transmitted) and a monochromatic (reflected)
branch. The first monochromator crystal with a thickness of $\simeq$~100~$\mu$m provides sufficient 
X-ray transmittance to enable simultaneous operation of two beamlines.
Here we report on the design, fabrication, and X-ray characterization of the
first and second (300-$\mu$m-thick) crystals utilized in the monochromator and the optical 
assemblies holding these crystals. 
Each crystal plate has a region of about 5~$\times$~2~mm$^2$ with low defect concentration, 
sufficient for use in X-ray optics at the LCLS. The optical assemblies holding the crystals were designed to provide 
mounting on a rigid substrate and to minimize mounting-induced crystal strain. 
%The thermal contact was due to the polished surface of the diamond crystal gently pressed against 
%the polished surface of the substrate. 
%To improve heat transfer and radiation hardness all parts of the optical assemblies were fabricated out of diamond materials. 
The induced strain was evaluated using double-crystal X-ray topography and was found to be small 
over the 5~$\times$~2~mm$^2$ working regions of the crystals.
\end{abstract}
%Use 

\maketitle

\section{Introduction}
 
Top-quality type IIa diamond crystals suitable for advanced applications in X-ray crystal optics 
at synchrotrons and X-ray free-electron lasers (XFELs) have recently become available due to refinements 
in the high-pressure high-temperature (HPHT) crystal synthesis method \cite{Burns09,Polyakov11,Sumiya12}. 
Diamond crystal plates with (111) surface orientation are of primary importance for front-end diffracting X-ray optics due 
to a greater intrinsic energy bandwidth of the 111 Bragg reflection and the resulting flux of the reflected X-rays compared
with those of higher-order reflections. 
However, production of the (111) crystal plates with large defect-free regions is a more challenging task 
compared with production of plates of other orientations close to the [001] direction. 
This is due to ($i$) the (001) diamond HPHT growth sector has the best crystal quality and the lowest impurity 
concentration; and ($ii$) the \{111\} crystal faces are the most resistant to polishing since these planes have 
the highest atomic density.

\begin{figure}[t!]
\setlength{\unitlength}{\textwidth}
\begin{picture}(0.5,0.3)(0,0)
\put(0.03,0.0){\includegraphics[width=0.4\textwidth]{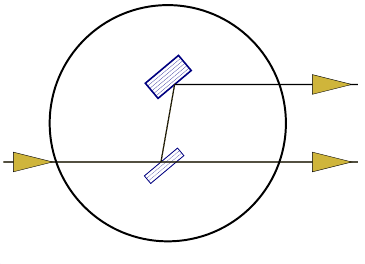}}
\end{picture}
\caption{Scheme of a double-crystal beam-multiplexing monochromator in Bragg reflection geometry. The first crystal is sufficiently thin to avoid substantial losses in the transmitted branch due to photoabsorption.}
\label{fig:dcm}
\end{figure}

The primary application of diamond (111) crystal plates in X-ray optics are the 
high-heat-load double-crystal monochromators at synchrotron undulator beamlines (e.g., \cite{Grubel96,Fernandez97,Yabashi07}).
A scheme of a double-crystal monochromator in Bragg reflection geometry is shown in Fig.~\ref{fig:dcm}.
If the first crystal is made sufficiently thin, the beam transmitted through it can be used 
in a parallel experiment downstream, i.e., beam multiplexing. 
Utilization of X-ray monochromators with low absorption crystals for beam multiplexing was pioneered 
at the TROIKA beamline at the ESRF (European Synchrotron Radiation Facility, Grenoble, France)~\cite{ANielsen94,Grubel94,Grubel96}. 
This beam multiplexing approach gains even more importance for hard X-ray free-electron lasers, 
which recently redefined the frontiers of X-ray sciences \cite{Emma10,Ishikawa12,Emma12}. 
Using a single straight electron trajectory of an XFEL, multi-user operations cannot be achieved 
as is commonly accomplished at storage-ring-based synchrotron sources. 
However, multiplexing can be performed by means of X-ray optics, which enables simultaneous 
delivery of portions of the XFEL beam to several experiments. This yields an increase in the total number of 
performed experiments and thus reduces the high XFEL operating cost per experiment.

Highly developed crystal fabrication and processing methods for silicon (Si) 
have led to a few attempts to use Si crystals for beam-multiplexing XFEL monochromators. 
Greater X-ray absorption in Si requires utilization of ultra-thin crystals (thicknesses $\approx$~5-10~$\mu$m)
to achieve sufficient transmittance of the XFEL beam (30-80\%) over a photon energy of 4-10~keV.
High-quality ultra-thin Si crystals for XFEL beam-multiplexing monochromators 
can be manufactured using state-of-the-art processing methods \cite{Feng12,Osaka13}; however, 
such ultra-thin crystals have been found unstable in the XFEL beam~\cite{Feng12,FZL13}. 

Similar levels of transmittance can be achieved with crystal thicknesses of 50-100~$\mu$m in the case of diamond due 
to its lower X-ray absorption. 
The greater crystal thickness and Young's modulus for diamond ensure a substantially greater stiffness of the sample.
Diamond crystals with such thicknesses are expected to be more stable in the XFEL beam if an appropriate crystal mounting
scheme is provided. 

Because of the small divergence ($\lesssim$~1~$\mu$rad rms) of the XFEL beam its wavefront is particularly sensitive to imperfections of X-ray optics. Disturbances of the wavefront introduced by the optics should be much less or at least comparable to the beam divergence, which is a challenging requirement, especially for diamond crystal optics prone to crystal defects.
Along with the presence of intrinsic defects, a mounting-induced crystal strain is another main factor that leads to deterioration of the crystal diffraction performance, which disturbs the radiation wavefront. 
The mounting solution previously developed by some of the co-authors for the diamond (001) self-seeding XFEL monochromator \cite{Stoupin_DRM13,Shu_2013jpcs1} was found to show limitations in heat transfer between the crystal plate and the graphite holder where the plate was loosely mounted \cite{Feng13}.

In this work we report on the production of 100-$\mu$m-thick and 300-$\mu$m-thick type IIa HPHT diamond crystal plates with (111) orientation. Each crystal plate has a region with low concentration of defects (working region) of about 5~$\times$~2~mm$^2$ sufficient for use in XFEL optics. More importantly, we present a solution for strain minimization as a special mounting of such crystals on a rigid substrate. All parts of the optical assembly were fabricated out of diamond, which improved heat transfer and radiation hardness of the device. The mounting-induced strain was evaluated using double-crystal X-ray topography in rocking curve diffraction imaging mode. The mounting induced crystal distortion (1.5 $\mu$rad rms variation in the rocking curve peak position across the working region) was substantially smaller than the full width at half maximum (FWHM) of the rocking curve width (25-30 $\mu$rad) and still comparable to the divergence of the XFEL beam. 
Two such assemblies were installed into the vacuum tank of the beam-multiplexing monochromator at the XPP instrument of the Linac Coherent Light Source and successfully tested \cite{Zhu13}. The monochromator was operated in Bragg reflection geometry, as shown in Fig.~\ref{fig:dcm}. The capability of splitting the XFEL beam into a pink (transmitted) and a monochromatic (reflected) branch was demonstrated, which enables the use of the XFEL beam in two experiments simultaneously.

\section{Diamond (111) crystal plates}

\begin{figure}
\setlength{\unitlength}{\textwidth}
\begin{picture}(0.5,0.76)(0,0)
\put(0.0,0.0){\includegraphics[width=0.48\textwidth]{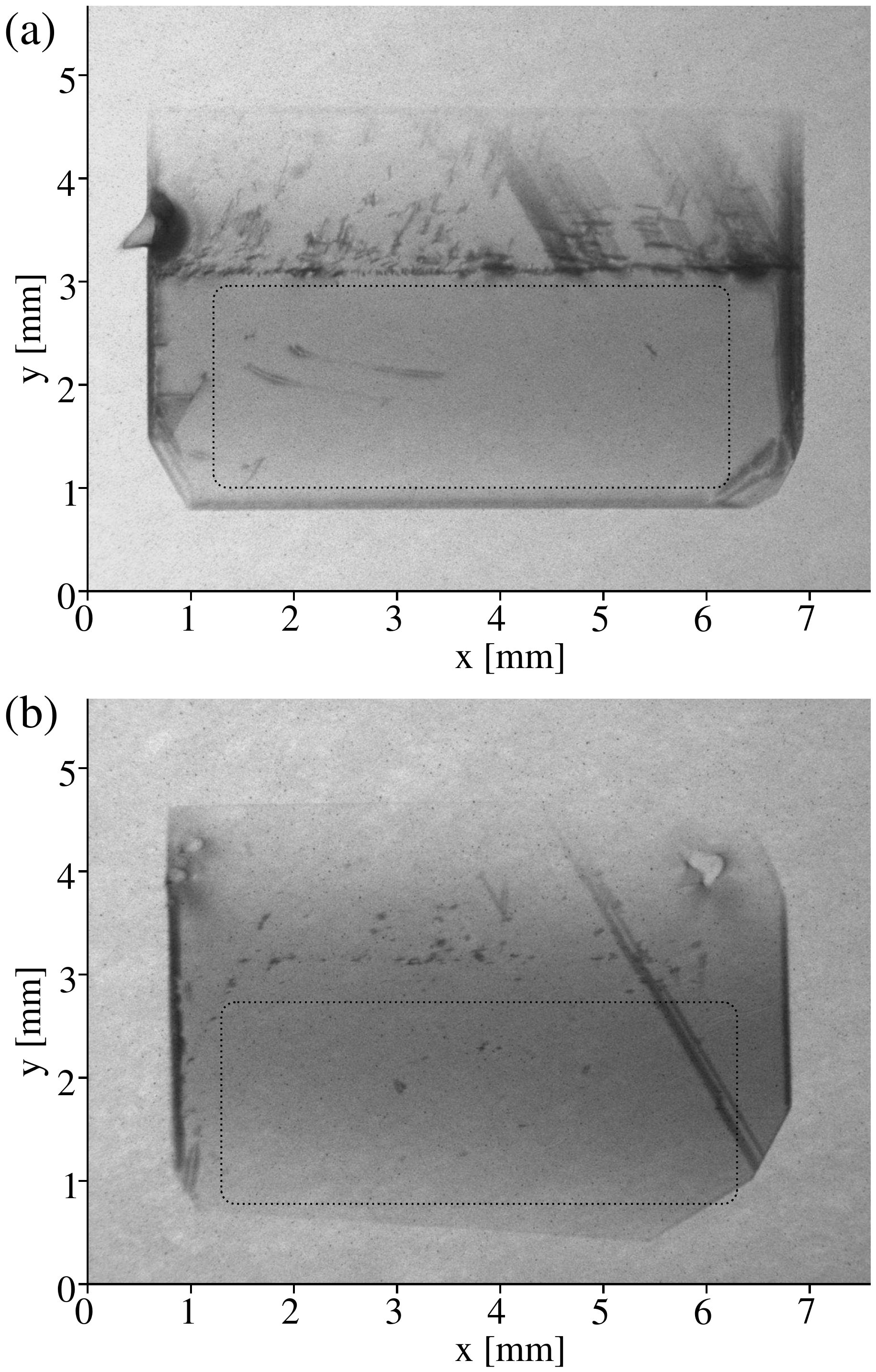}}
\end{picture}
\caption{White-beam X-ray topographs of the selected diamond (111) crystal plates:
(a) 300-$\mu$m-thick plate (second monochromator crystal) and (b) 
100-$\mu$m-thick plate (first monochromator crystal). The images are obtained
from $3\bar{1}\bar{1}$ diamond reflection in Laue geometry. The dotted box illustrates the 
5~$\times$~2~mm$^2$ working regions containing only a few defects.}
\label{fig:wbt}
\end{figure}

Type IIa diamond single crystals were grown at the Technological Institute for Superhard and Novel Carbon Materials (TISNCM) using the temperature-gradient method at high static pressure and high temperature (e.g., \cite{Blank07,Polyakov11}). The temperature of crystallization was 1460 $^{\circ}$C at a pressure of 5.5 GPa. After the crystallization process, diamond crystals were cut by a laser along the $\{111\}$ crystal plane with a $\simeq$~2$^{\circ}$ angular offset from the plane. 
The angular offset was introduced to facilitate mechanical polishing of the crystal plates. 
The plates were polished to a micro-roughness of $\approx$~10~nm (rms). 

Preliminary selection of crystal plates with low density of crystal defects was made using Lang X-ray topography. 
For the final selection of diamond crystal plates, white-beam X-ray topography was performed at the MRCAT 10BM (bending magnet) beamline~\cite{Kropf_AIP10} of the Advanced Photon Source. 
White-beam X-ray topographs were obtained in Laue (transmission) geometry on the diamond plates enclosed in Kapton film holders.
The source-to-sample distance was $L\simeq$~27~m, the effective source size at the beamline was $s \approx 300$~$\mu$m, 
and the distance from the sample to the observation plane was $d \simeq$~0.1~m. Thus, the expected spatial resolution in the observation plane was $\delta = sd/L \simeq$~1~$\mu$m, which was comparable with the photographic resolution of the utilized X-ray film (AGFA STRUCTURIX D3-SC). 

We note that the image resolution was somewhat reduced due to crystal instability caused by heat load of the X-rays during the shortest available exposure time ($\approx$~2~s). Multiple images were taken to mitigate this problem. High-quality topographs were obtained from the $3\bar{1}\bar{1}$ reflections. They are shown in Fig.~\ref{fig:wbt}(a) and (b) for the selected 300-$\mu$m-thick and 100-$\mu$m-thick crystal plates, respectively.

The dotted box in  Fig.~\ref{fig:wbt}(a) and (b) shows the 5~$\times$~2~mm$^2$ working regions originated for the most part from the 
(001) growth sector with a low concentration of defects. A few defects of about 10~$\mu$m in size are noticeable in the working regions of the crystals. These can be attributed to minor dislocations and micro-inclusions. The diagonal lines represent stacking faults that likely originate at the growth sector boundaries. Nevertheless, the crystals are of remarkable quality in the  5~$\times$~2~mm$^2$ region, especially if compared to commercially available type IIa diamond (111) crystals (e.g., the crystal studied earlier by some of the authors \cite{Shvyd'ko09}).   

\section{Crystal mounting method}

\begin{figure}[!t]
%\centering\includegraphics[width=0.45\textwidth]{design2.pdf}
\setlength{\unitlength}{\textwidth}
\begin{picture}(0.5,0.35)(0,0)
\put(0.0,0.0){\includegraphics[width=0.45\textwidth]{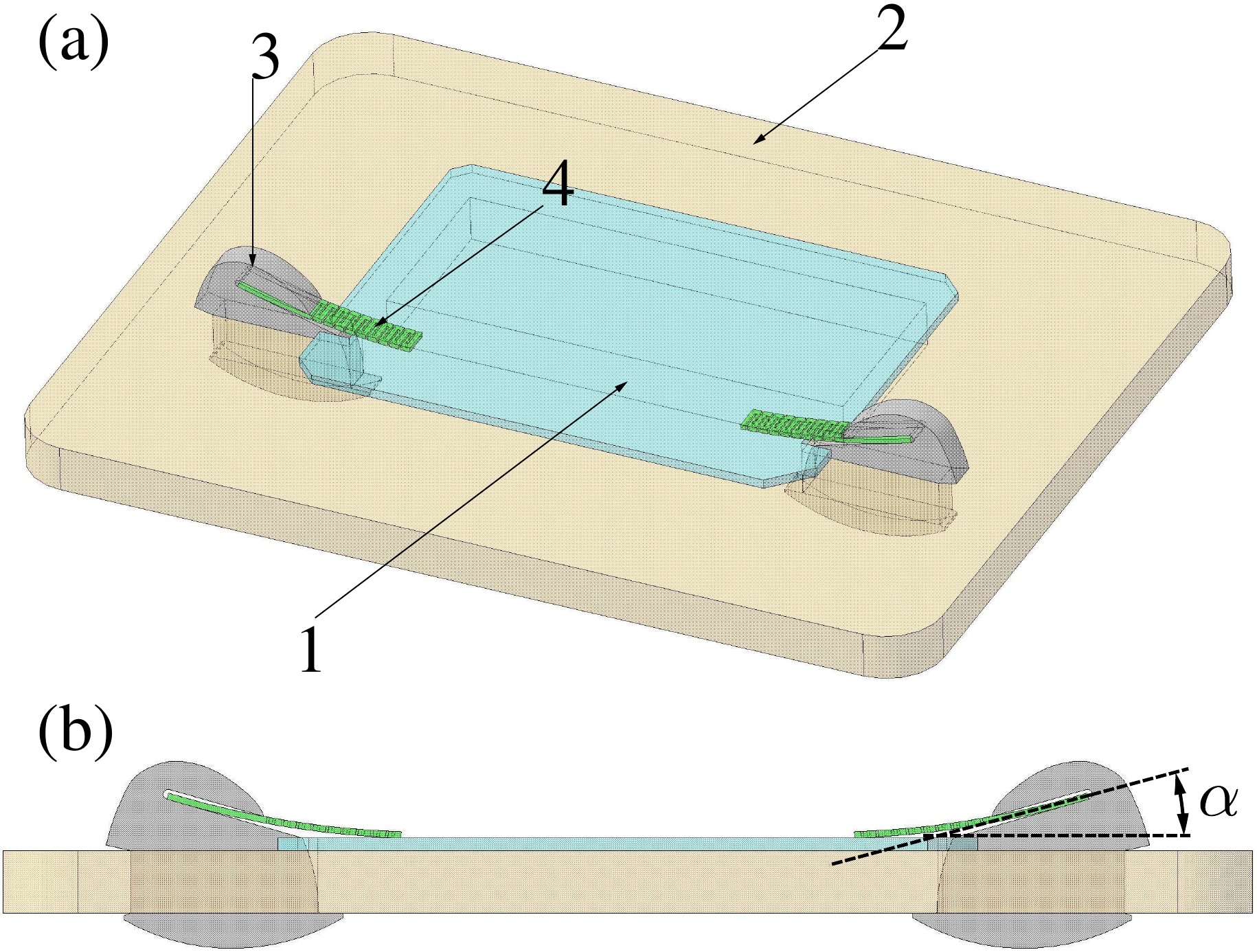}}
\end{picture}
\caption{Scheme of the all-diamond optical assembly.
(a) Perspective view showing the components of the assembly: type IIa HPHT diamond (111) crystal plate 1, 
a CVD diamond substrate 2, diamond restrainers 3, and perforated CVD diamond springs 4.
(b) Side view shows the orientation of the grooves in the restrainers for insertion of the CVD diamond spring. 
The groove is at an angle $\alpha$ relative to the surface of the optical element.}
\label{fig:design}
\end{figure}

\begin{figure}
%\centering\includegraphics[width=0.45\textwidth]{CVDspring.pdf}
\setlength{\unitlength}{\textwidth}
\begin{picture}(0.5,0.35)(0,0)
\put(0.0,0.0){\includegraphics[width=0.45\textwidth]{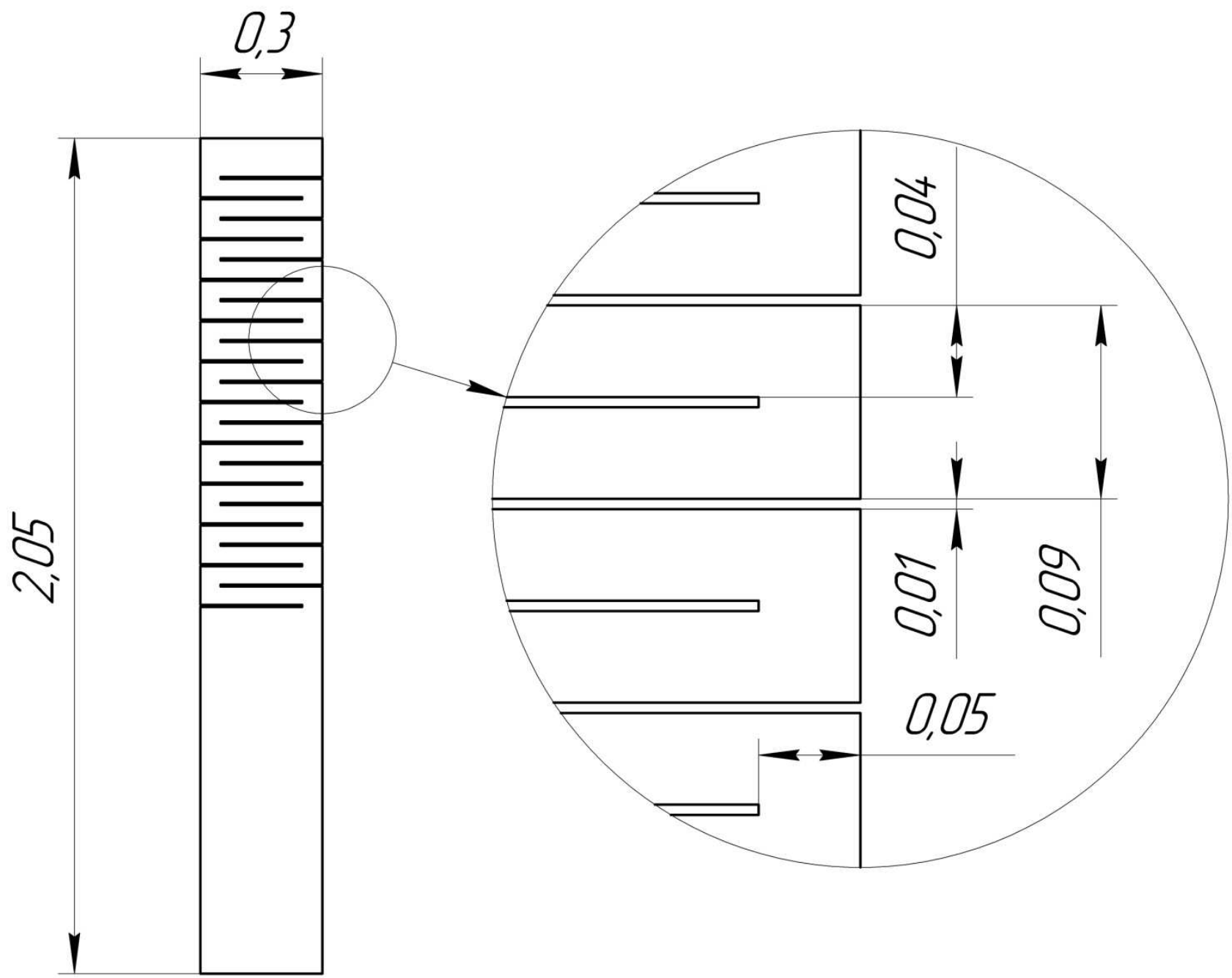}}
\end{picture}
\caption{Scheme of the CVD spring. The inset is an enlarged view showing perforation dimensions. All dimensions are given in mm.}
\label{fig:spring}
\end{figure}

\begin{figure*}[!t]
\setlength{\unitlength}{\textwidth}
\begin{picture}(1.0,0.35)(0,0)
\centering\includegraphics[width=1.0\textwidth]{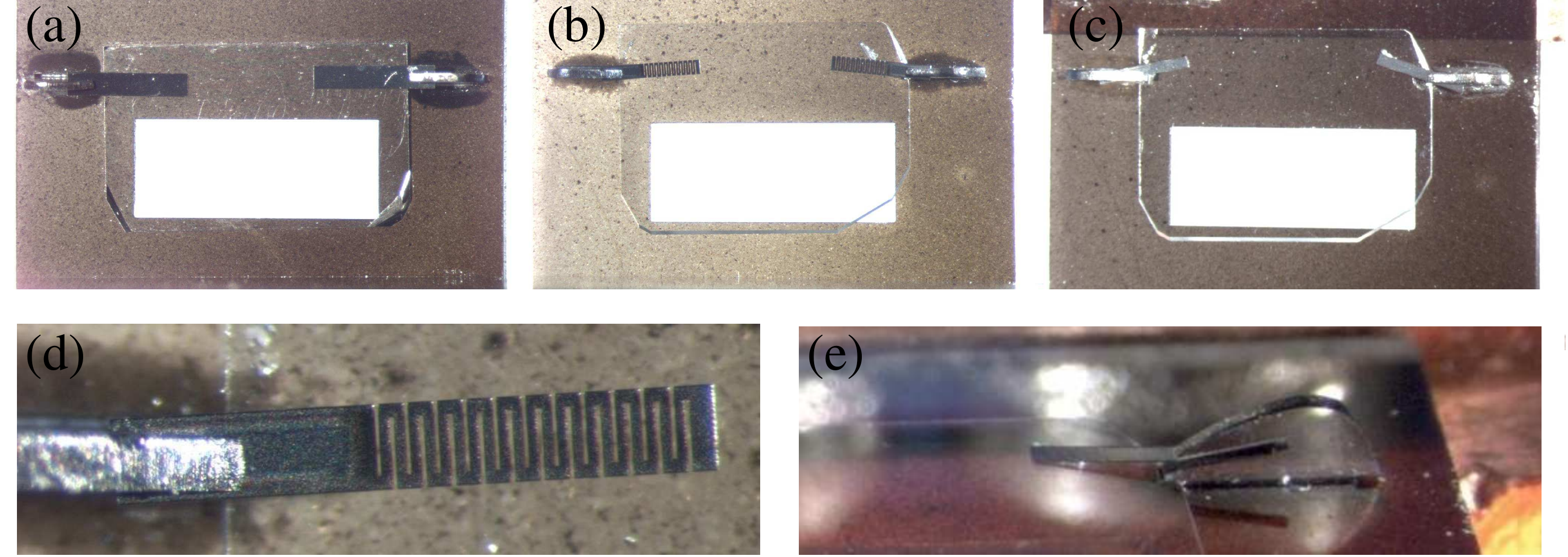}
\end{picture}
\caption{All-diamond assemblies (top view) with high-quality diamond crystal plates (type IIa, HPHT, (111) surface orientation) mounted on a CVD diamond substrate: (a) assembly with a 300-$\mu$m-thick diamond crystal plate using non-perforated CVD diamond springs; (b) assembly with a 100-$\mu$m-thick diamond crystal plate using perforated CVD diamond springs; (c) assembly with the same 100-$\mu$m-thick diamond crystal plate using non-perforated CVD diamond springs.  
The bright rectangles in (a-c) are the 5~$\times$~2~mm$^2$ windows in the CVD substrates for passage of X-rays transmitted through the diamond plates. Selected microscope images: (d) an enlarged view of the perforated CVD diamond spring, and (e) a side perspective view of a diamond restrainer. A CVD spring that acts on the HPHT crystal is inserted into a groove in the restrainer.
}
\label{fig:asmb}
\end{figure*} 

The relative intrinsic energy bandwidth of X-ray Bragg reflection is a very small quantity 
($\Delta E/E \approx 10^{-5} - 10^{-7}$), which is related to the small intrinsic angular width 
$\Delta \theta$~$\approx$~$1 - 100$~$\mu$rad (except Bragg-backscattering cases $\theta \simeq$~90$^{\circ}$):
\begin{equation}
\frac{\Delta E}{E} = \frac{\Delta \theta}{\tan{\theta}} .
\end{equation}

Since the relative energy bandwidth is small, a similarly small relative crystal strain $\delta a / a \sim \Delta E /E$ ($a$ is the crystal lattice parameter) can distort the shape of the intrinsic curves and reduce performance of the crystal as a monochromator. 
Thin Bragg crystals are particularly affected since relative deformations of the crystal are larger for a given applied stress (force).
A simple fact that a thin crystal has a fixed contact point with a holder (achieved via bonding using glue, tape, or even grease) can be distinguished in Bragg diffraction from a situation where the crystal has a nearly free boundary (i.e.,  motion of the crystal is constrained only by friction against a crystal holder). Such a nearly free boundary condition is appropriate for certain applications; however, in this case the crystal is very susceptible to vibrations. Also, thermal transfer between the crystal and the holder could be limited due to reduced surface contact. These factors may severely deteriorate diffraction performance of the crystal under an intense incident X-ray beam.

In this work, an alternative approach to mounting a diffracting crystal is explored, which includes a gentle pressure (a spring force) applied to the crystal placed on a rigid substrate. This force is applied in a controlled manner, such that the resulting strain is evaluated and limited to desired specifications. 
The optical assembly is schematically shown in Fig.~\ref{fig:design}. In order to minimize issues related to differential thermal expansion, and to improve radiation hardness and thermal transfer all parts of the assembly were manufactured out of diamond materials. To improve thermal transfer between the substrate and the crystal their surfaces in contact were polished. 

A high-quality type IIa HPHT diamond (111) crystal plate is mounted on a substrate fabricated out of polycrystalline chemical vapor deposited (CVD) diamond. 
The substrate is sufficiently thick (500~$\mu$m) to provide rigid support for the diamond plate. The surface of the substrate in contact with the diamond plate was polished to $\approx$~10~nm (rms) micro-roughness. The substrate has two small rectangular openings for insertion of restrainers made of low-quality type IIa HPHT diamond and one large rectangular window (5~$\times$2~mm$^2$) for passage of X-rays transmitted through the diamond plate. The diamond plate has two small rectangular cuts on the sides such that the restrainers prevent lateral displacement of the plate on the substrate.
The restrainers have grooves for insertion of 15-$\mu$m-thick CVD diamond stripes (CVD springs) that provide gentle force on the diamond plate, thus preventing its vertical displacement. The force acting on the diamond is 
\begin{equation}
F \propto \frac{\alpha }{L^2}k,
\label{eq:force}
\end{equation}
where $\alpha$ is the bending angle of the CVD spring as shown in Fig.~\ref{fig:design}(b), $L$ is the 
length of the spring, and k is the flexural stiffness of the spring. 
The spring can be approximated by a cantilever beam with an area moment of inertia $I$, such that $k = EI$, where $E$ is the Young's modulus. Thus, for a fixed angle $\alpha$ (in our design $\alpha \simeq 15^{\circ}$) the force can be controlled by either variation of the length of the spring or variation of the spring stiffness.

Increasing the length of the spring was found to be insufficient in our experiments to reach very small forces to minimize mounting strain in diamond crystal plates with $\approx$~100-$\mu$m thickness.
An additional spring force reduction was achieved by changing the stiffness of the CVD springs. The best solution to substantially reduce the stiffness of the springs was to introduce periodic strain-relief cuts (perforation) in the spring as shown in Fig.~\ref{fig:spring}.

All diamond components were manufactured at TISNCM using HPHT and CVD diamond synthesis methods and a controlled precision laser cutting of diamond materials. 
Polycrystalline CVD diamond films were deposited onto 640-$\mu$m-thick Si single-crystal substrates. Prior to deposition the Si substrates were ultrasonically treated in diamond powder suspension in ethanol. Diamond deposition process was carried out in a bell jar-type Microwave Plasma-Assisted Chemical Vapor Deposition (MWPA CVD) reactor. 
The substrate temperature was maintained at 860-870 C$^\circ$. The methane concentration was 6\% in a total flow of 500 sccm. 
The microwave power was in the range 3 - 3.2 kW at a total pressure of about 200 mbar during deposition of thick films for production of the 500-$\mu$m-thick CVD diamond substrates, and 130 mbar during deposition of thin flexible films for production of the 15-$\mu$m-thick CVD springs. After the deposition the diamond films were separated from the silicon substrates by dissolution of the substrates 
in 44 wt.\% solution of KOH in water at 50$^\circ$. 

The devices were assembled and cleaned at the Advanced Photon Source. Microscope images of the all-diamond assemblies are shown in Fig.~\ref{fig:asmb}.
Figure~\ref{fig:asmb}(a) shows an assembly with a 300-$\mu$m-thick diamond crystal plate mounted using non-perforated CVD diamond springs. 
Figure~\ref{fig:asmb}(b) shows an assembly with a 100-$\mu$m-thick diamond crystal plate mounted using perforated CVD diamond springs.
Figure~\ref{fig:asmb}(c) shows an assembly with the same 100-$\mu$m-thick diamond crystal plate mounted using non-perforated CVD diamond springs. 
Figure~\ref{fig:asmb}(d) shows an enlarged view of the perforated spring acting on the 100-$\mu$m-thick diamond crystal plate, and Fig.~\ref{fig:asmb}(e) shows an enlarged side perspective view of a restrainer with an inserted CVD spring. 

%%%%%%%%%%%%%%%%%%%%%%%%%%%%%%%%%%%%%%%%%%%%%%%%%%%%%%%%%%%%%%%%%%%%%%%%%%%%%%%%%%%%%%%%%%%%%%%%%%%%%%%%%%%%%%%%%%%%%%%%%%%%%%%%%%%%%%%%%%%%%%%%%%%%%%%%%%%   
\section{Characterization of mounting-induced strain}
\subsection{Experimental}

\begin{figure}[h!]
\setlength{\unitlength}{\textwidth}
\begin{picture}(0.5,0.22)(0,0)
\put(0.01,0.0){\includegraphics[width=0.45\textwidth]{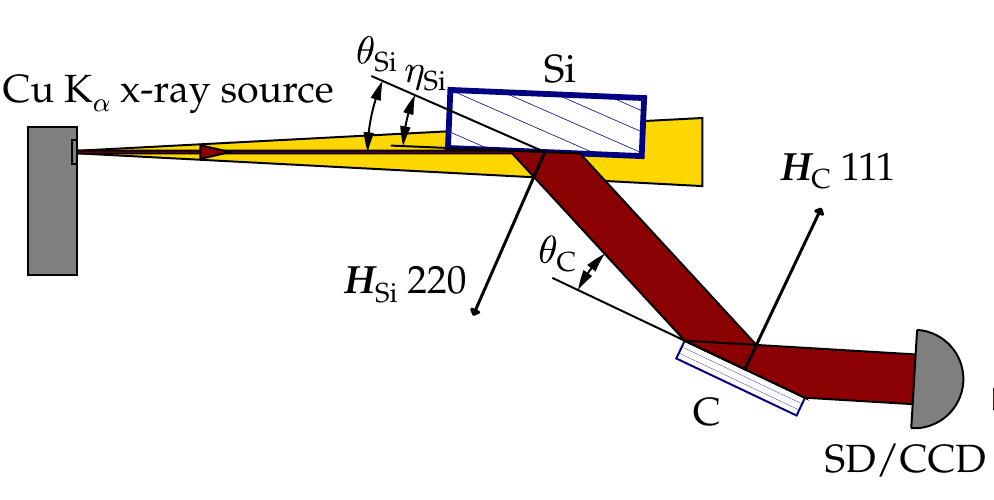}}
\end{picture}
\caption{Double-crystal topography setup for mapping the rocking curve of the 111 diamond reflection (see text for details).}
\label{fig:setup}
\end{figure}

Rocking curve measurements and rocking curve imaging~\cite{Lubbert00} were the main diagnostic methods 
for evaluating the mounting-induced crystal strain in the all-diamond assemblies.
Double-crystal topography using a Cu K$\alpha$ rotating anode 
X-ray source was performed to map the rocking curve of diamond crystal plates mounted in the holder assemblies. 
A Si (220) first crystal with an asymmetry angle $\eta_{\mathrm{Si}} \simeq$~22.2$^{\circ}$ 
was used to collimate an X-ray beam incident on the diamond crystal plate (C) mounted in the all-diamond assembly (Fig.~\ref{fig:setup}). 
The diamond crystal was set for the 111 reflection. The Bragg angle of the diamond crystal 
was $\theta_{\mathrm{C}} \simeq$~22.0$^{\circ}$, which was close to the Bragg angle of the collimator 
crystal $\theta_{\mathrm{Si}} \simeq$~23.7$^{\circ}$. 

In order to align the diamond crystal for the 111 reflection and to obtain the rocking curve of the entire crystal (total rocking curve),
the intensity reflected from a uniformly illuminated diamond crystal was measured using a scintillation detector (SD).
A CCD camera (with a resolution $60\times60$~$\mu$m$^2$) replacing SD was used to obtain a series of X-ray diffraction images at different angular 
positions of the diamond crystal through the rocking curve of the 111 reflection. The image data were sorted such that a local rocking curve was obtained  
for every pixel, thus making it possible to map the rocking curve parameters over the entire crystal (i.e., rocking curve imaging). 

The angular resolution in the rocking curve scan was $\approx$~4~$\mu$rad, which is small compared with the FWHM 
of the ideal rocking curve in the double-crystal geometry ($25$~$\mu$rad). The latter 
was evaluated using the dynamical theory of X-ray diffraction assuming a uniform angular distribution of the radiation incident 
on the first crystal (over an angular range exceeding the angular acceptance region of the crystal) 
and a photon energy distribution of the Cu K$\alpha$ characteristic lines \cite{Hartwig93}.

\subsection{Analysis of total and local rocking curves}

\begin{figure*}
\setlength{\unitlength}{\textwidth}
\begin{picture}(1.0,0.35)(0,0)
\put(0.0,0.0){\includegraphics[width=1.0\textwidth]{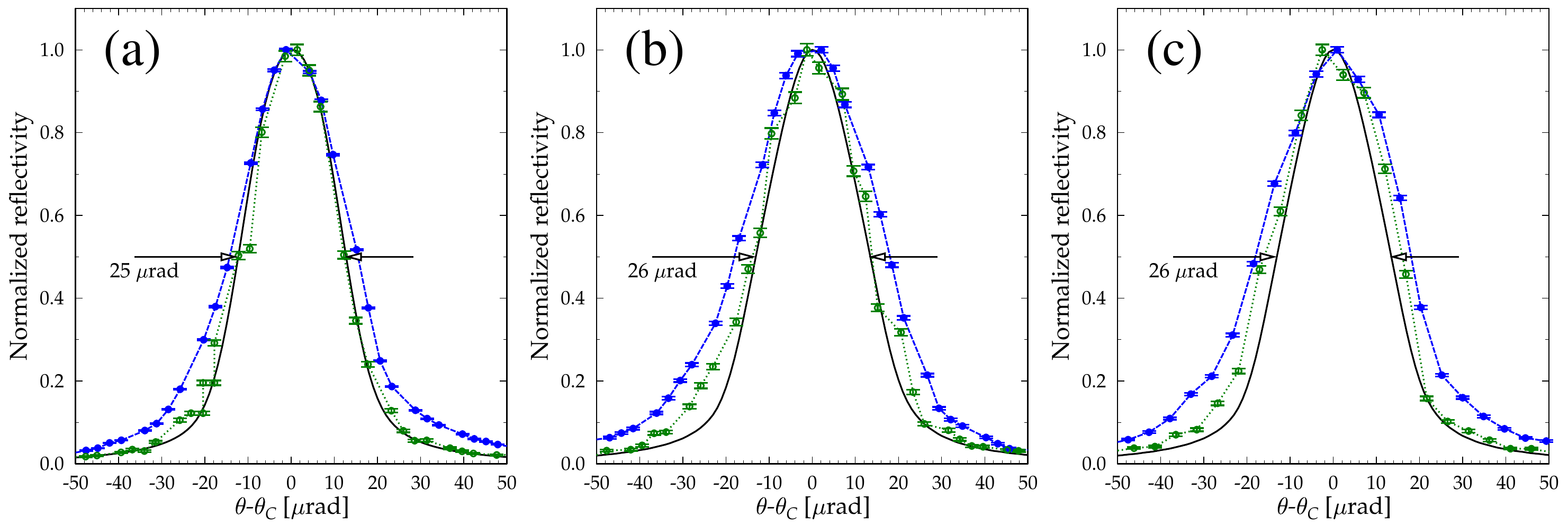}}
\end{picture}
\caption{Rocking curves of diamond crystal plates in the all-diamond assemblies: total (filled blue circles, dashed blue line), local (open green circles, dotted green line), and theoretical (solid black line) for (a) a 300-$\mu$m-thick plate mounted using non-perforated CVD diamond springs; (b) a 100-$\mu$m-thick plate mounted using perforated CVD diamond springs; and (c) a 100-$\mu$m-thick plate mounted using non-perforated CVD diamond springs.}
\label{fig:rc_abc}
\end{figure*} 

Figure~\ref{fig:rc_abc} shows measured total rocking curves (blue circles and lines), the theoretical rocking curve (solid black line), and local rocking curves (green circles and lines) from a selected pixel at the center of the working region on the diamond (111) crystal plate. Figure~\ref{fig:rc_abc}(a) shows rocking curves for the 300-$\mu$m-thick crystal mounted using non-perforated CVD diamond springs (as shown in Fig.~\ref{fig:asmb}(a)). The FWHM of the local rocking curve matches that of the theoretical curve (25 $\mu$rad) while the total curve exhibits an additional broadening. This observation suggests that the broadening of the total rocking curve is due to the presence of defects in the crystal (i.e., intrinsic strain). The defects are located outside the working region of the crystal as will be shown further. 

Figure~\ref{fig:rc_abc}(b) similarly shows rocking curves for the 100-$\mu$m-thick crystal mounted using perforated CVD 
diamond springs (as shown in Fig.~\ref{fig:asmb}(b)). In this case, both the total rocking curve and the local rocking curve exhibit an additional broadening compared to the FWHM of the theoretical rocking curve for the 100-$\mu$m-thick crystal (26 $\mu$rad).

Figure~\ref{fig:rc_abc}(c) shows rocking curves for the 100-$\mu$m-thick crystal mounted using non-perforated CVD diamond springs.
The broadening of the local rocking curve becomes somewhat larger, which suggests that the use of perforated springs is preferable 
for minimization of strain. However, this observation needs an additional verification, since there could be a variation in the local 
rocking curve from one pixel to another. 

The values of FWHM of the measured total and local rocking curves and those of the theoretical rocking curves are given in Table~\ref{tab:fwhm} for each case. The FWHM of the total rocking curve (36 $\mu$rad) is about the same for (b) and (c), 
which indicates that the broadening is dominated by crystal defects and that this parameter is not very sensitive to mounting strain. 
The broadening is smaller than the intrinsic angular width (Darwin width) of the reflection ($\simeq$~23.1~$\mu$rad). 
For certain applications of diamond in X-ray optics this validates the quality of the entire crystal and the mounting method using either springs. However, in our case the desired level of broadening should be less than or comparable with the angular divergence of the XFEL beam (1-2~$\mu$rad rms) to minimize disturbance of the radiation wavefront. Therefore, a more detailed characterization of the working crystal region is required. 
\begin{table}[!h]
\centering                    % centering table
\caption{Total ($\Delta \theta_{tot}$), local ($\Delta \theta$), and theoretical ($\Delta \theta_0$) rocking curve FWHM for assemblies (a), (b), and (c) (Fig.~\ref{fig:asmb}) with different diamond (111) crystal plate thicknesses ($t$) and CVD springs.}
\begin{tabular}{l c c c c c}
\hline\hline
assembly  & $t$        & CVD springs     &$\Delta \theta_{tot}$  &$\Delta \theta$  & $\Delta \theta_0$  \\
          & ($\mu$m)   &                 &($\mu$rad)             &($\mu$rad)             &($\mu$rad)            \\
\hline
(a)      & 300         & non-perforated  &30                     &25                     &25                    \\
(b)      & 100         & perforated      &36                     &28                     &26                    \\         
(c)      & 100         & non-perforated  &36                     &32                     &26                    \\ 
\hline
\end{tabular}
\label{tab:fwhm}
\end{table}

\subsection{Analysis of the rocking curve topographs}

\begin{figure*}[!t]
\setlength{\unitlength}{\textwidth}
\begin{picture}(1.0,0.45)(0,0)
\centering\includegraphics[width=1.0\textwidth]{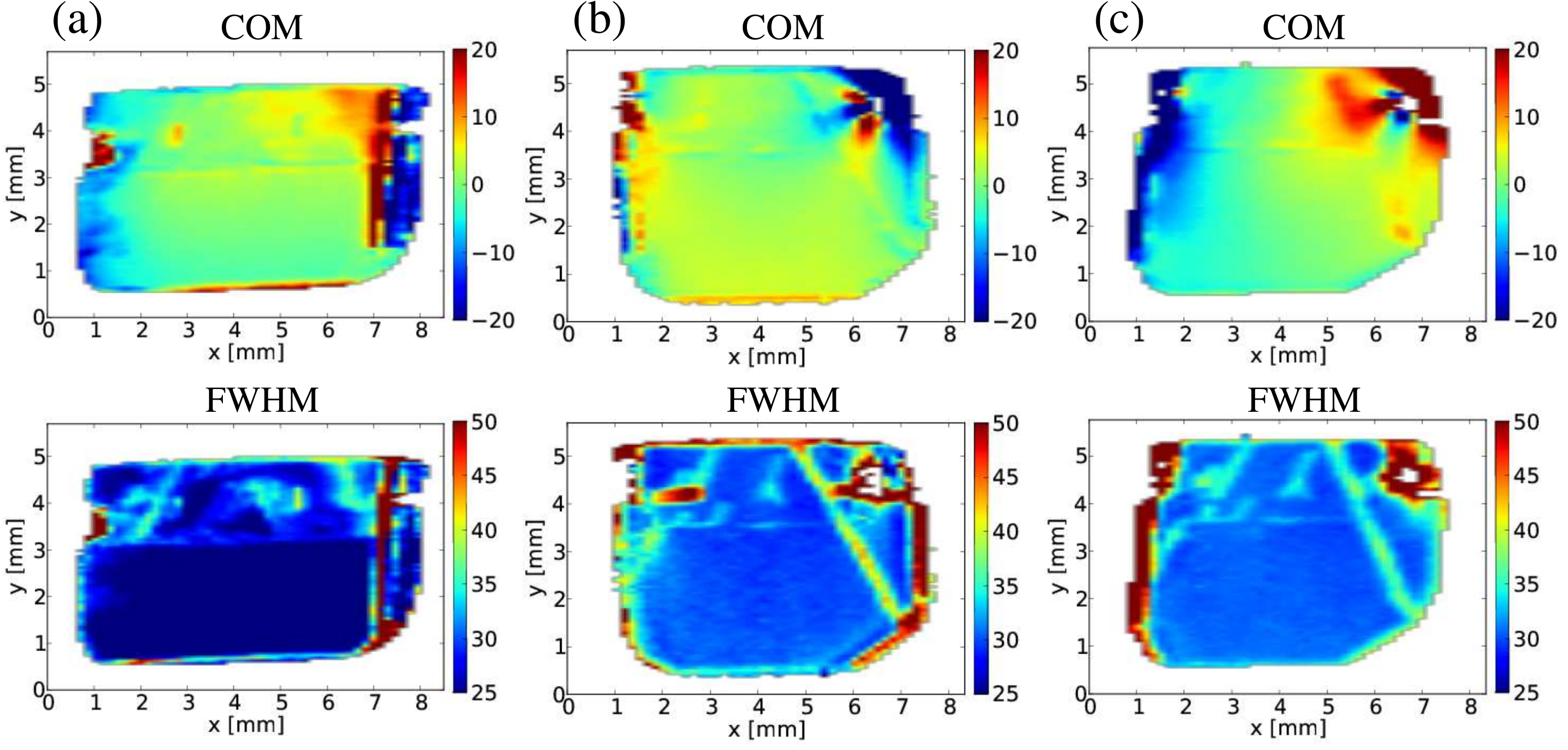}
\end{picture}
\caption{Double-crystal topographs showing maps of the rocking curve COM and the FWHM for:
(a) the 300-$\mu$m-thick crystal mounted using non-perforated CVD diamond springs; 
(b) the 100-$\mu$m-thick crystal mounted using perforated CVD diamond springs; and 
(c) the 100-$\mu$m-thick crystal mounted using non-perforated CVD diamond springs.
The units on the colorbars are given in $\mu$rad.}
\label{fig:maps1e}
\end{figure*} 

\begin{figure*}
\setlength{\unitlength}{\textwidth}
\begin{picture}(1.0,0.45)(0,0)
\centering\includegraphics[width=1.0\textwidth]{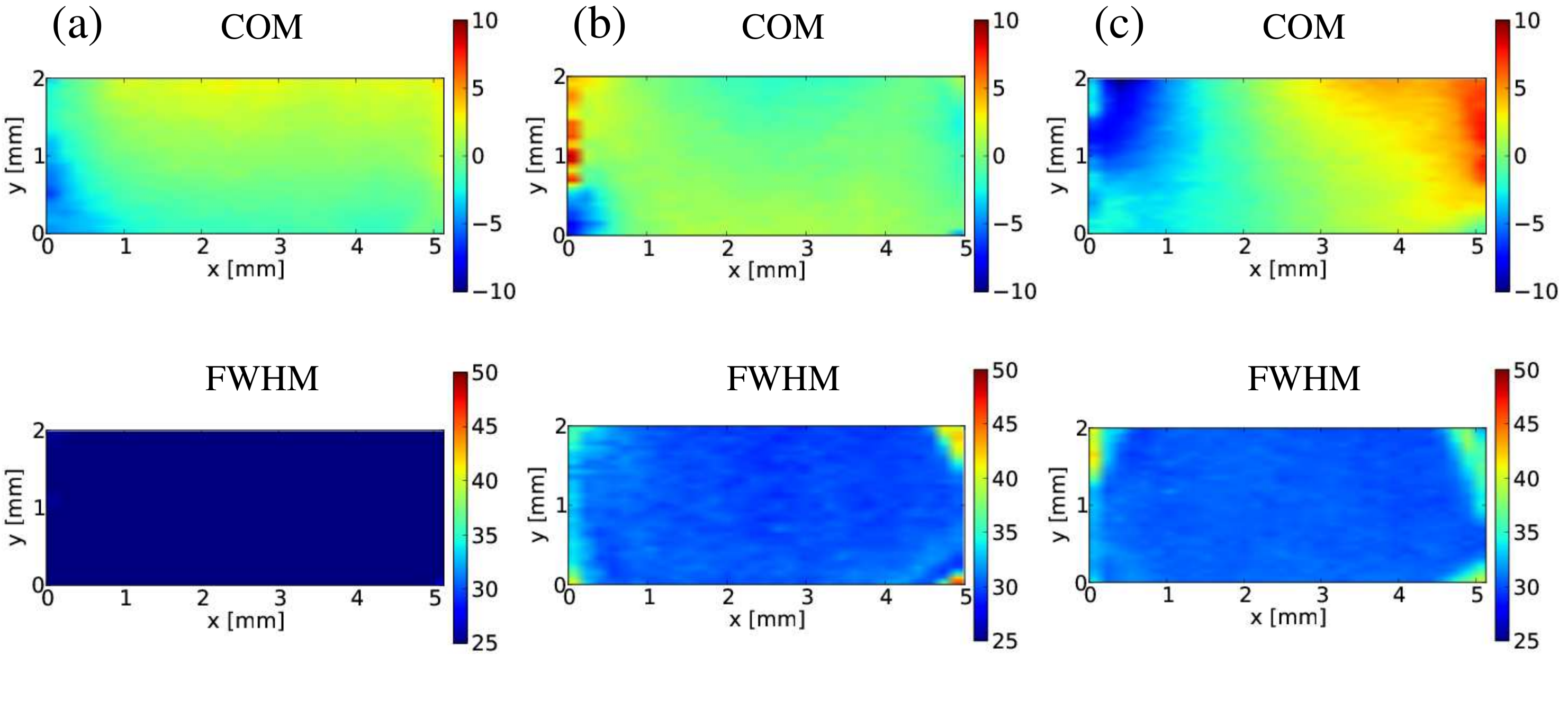}
\end{picture}
\caption{Double-crystal topographs (COM and FWHM) of the working crystal regions ($\simeq$~5$\times$~2~mm$^2$) for:
(a) the 300-$\mu$m-thick crystal mounted using non-perforated CVD diamond springs; 
(b) the 100-$\mu$m-thick crystal mounted using perforated CVD diamond springs; and 
(c) the 100-$\mu$m-thick crystal mounted using non-perforated CVD diamond springs.
The units on the colorbars are given in $\mu$rad.}
\label{fig:maps1e_wkr}
\end{figure*} 

The comparative analysis of the widths of the total and local rocking curves does not show a complete picture since it does not take into account a distribution of the rocking curve parameters over the crystal plate. A more sensitive indicator of the crystal strain is a local change in the lattice parameter or a tilt of the Bragg plane, which lead to a shift in the peak position of the local rocking curve. In our diffraction imaging analysis this shift is represented by the center of mass of the local rocking curve (COM). 

Figure~\ref{fig:maps1e} shows maps (topographs) of the rocking curve COM and FWHM for the 300-$\mu$m-thick plate mounted using the non-perforated springs (a), the 100-$\mu$m-thick plate mounted using the perforated springs (b), and the 100-$\mu$m-thick plate mounted using the non-perforated springs (c). 
These maps resemble the white-beam topographs given above. In particular, higher crystal quality in the working region and 
the diagonal lines (note higher contrast in FWHM) representing growth sector boundaries are clearly observed. 
The COM topographs indicate that a substantial strain is present in the upper part of the crystals. In the COM topographs for (a) and (b) the contrast is predominantly due to the intrinsic strain (presence of defects). In the case (c) the COM gradient is stronger, which indicates the presence of mounting-induced strain. 

The topographs of the working region are shown separately in Fig.~\ref{fig:maps1e_wkr}. The distribution of the FWHM is fairly uniform across the working region in each case. The COM topographs for (a) and (b) are again quite uniform, while in the case (c) a distinct gradient is observed. The topographs of the working region illustrate that for the 300-$\mu$m-thick plate mounting it is acceptable to use non-perforated springs (Fig.~\ref{fig:maps1e_wkr}(a)). 
However, for the 100-$\mu$m-thick plate mounting use of the non-perforated springs results in a higher level of strain (Fig.~\ref{fig:maps1e_wkr}(c)) compared to the case of perforated springs  (Fig.~\ref{fig:maps1e_wkr}(b)).

A simple statistical analysis of the rocking curve parameters was performed across the working region in each case. 
The values of standard deviation of the COM, standard deviation of the FWHM, and the average value for the FWHM are given
in Table~\ref{tab:wkrstat}. 
The largest variation of the local rocking curve peak position (COM) was for the 100-$\mu$m-thick plate mounted using non-perforated CVD springs (3.5 $\mu$rad). This value is still small compared to the average FWHM (31~$\mu$rad), which ensures a reasonable performance of the crystal in Bragg diffraction. The use of the perforated diamond CVD springs yields further minimization of the mounting-induced strain as reflected by the reduced standard deviation of the COM of the local rocking curve (1.5~$\mu$rad).

%\newpage
\begin{table}
\centering                    % centering table
\caption{Rocking curve statistical parameters across the working region for assemblies (a) (b) and (c) (Fig.~\ref{fig:asmb}): \\
$\delta \theta_{\rm COM}$ - standard deviation of the local rocking curve COM, \\
$\delta (\Delta \theta)$ -  standard deviation of the local rocking curve FWHM, \\
$<\Delta \theta>$ - average value of the local rocking curve FWHM.}
\begin{tabular}{l c c c c c c c}
\hline\hline
assembly  & $t$          & CVD springs     & $\delta \theta_{\rm COM}$   & $\delta(\Delta \theta)$   & $<\Delta \theta>$   \\
          & ($\mu$m)     &                 &($\mu$rad)         & ($\mu$rad)                & ($\mu$rad)  \\
\hline
(a)      & 300           & non-perforated  &1.4                &0.4                        &25           \\
(b)      & 100           & perforated      &1.4                &1.5                        &31           \\         
(c)      & 100           & non-perforated  &3.5                &1.5                        &31           \\ 
\hline
\end{tabular}
\label{tab:wkrstat}
\end{table}

\section{Conclusions}

In summary, all-diamond X-ray optical assemblies holding type IIa diamond (111) crystal plates were fabricated 
for the beam-multiplexing diamond monochromator at the LCLS. It is demonstrated how requirements on crystal quality and crystal mounting 
lead to an advanced application of diamond materials in XFEL optics. Two crystal plates of 100 $\mu$m and 300 $\mu$m thicknesses with low concentration of defects in a region of 5$\times$2~mm$^2$ were selected using white-beam X-ray topography. 
A dedicated crystal mounting method was developed for minimization of mounting strain. 
The mounting strain was evaluated using double-crystal X-ray topography in the rocking curve imaging mode.
The variation of the rocking curve peak position over the working region of the crystal plates was found to be substantially smaller than the angular width of the curve and comparable with the angular divergence of the XFEL beam.
With these assemblies installed in the beam-multiplexing monochromator at the LCLS, the capability of splitting the XFEL beam into a pink and a monochromatic branch was demonstrated \cite{Zhu13}. It was found that the measured bandwidth and throughput of the monochromator closely match the theoretical values, and the resulting beam profiles are only minimally disturbed.

\section{Acknowledgments}
B. Stephenson and L. Young are acknowledged for their support and interest in this work.
The help of our colleagues J. Maj, X. Huang, I. Lemesh, and S. Marathe is greatly appreciated. 
V. Srinivasan is acknowledged for engineering support of the project at the LCLS.
The present work was supported through a research grant from the Russian Ministry of Education and Science (Contract Nos.16.552.11.7014). 
Use of the Advanced Photon Source was supported by the U. S. Department of Energy, Office of Science, under Contract No. DE-AC02-06CH11357.  MRCAT operations are supported by the U.S. Department of Energy and the MRCAT member institutions.

%\bibliography{/home/oxygen/SSTOUPIN/Work2/xfelo/bib/mybib/references}% Produces the bibliography via BibTeX.
%\bibliography{/home/sstoupin/Work2/xfelo/bib/mybib/references}
%\bibliographystyle{spiebib} 

\end{document}